\input{aipcheck}

\documentclass[
    ,final            
  ]
  {aipproc}
\layoutstyle{6x9}
\usepackage{amsmath}
\usepackage{subfig}

\begin{document}

\title{Electroproduction of $\Lambda$(1405)}


\author{Haiyun Lu}{
  address={Carnegie Mellon University, Pittsburgh, Pennsylvania 15213, USA}
}

\author{Reinhard Schumacher}{
  address={Carnegie Mellon University, Pittsburgh, Pennsylvania 15213, USA}
}

\author{Brian Raue}{
  address={Florida International University, Miami, Florida 33199, USA}
}
\author{Marianna Gabrielyan}{
  address={Florida International University, Miami, Florida 33199, USA}
}

\begin{abstract}
The electroproduction of $\mathrm{K}^+ \Lambda\mathrm{ (1405)}$ was studied by analyzing the E1F data set collected in Hall B at Jefferson Lab. The analysis utilized the decay channel $\Sigma^+ \pi^-$ of the $\Lambda$(1405) and $p \pi^0$ of the $\Sigma^+$. Simulations of background, $\Lambda$(1405) and $\Lambda$(1520) production according to PDG values were performed by using standard CLAS analysis tools adapted for the E1F run. Fits of the acceptance-corrected simulations were made to the acceptance-corrected data to determine contributions from signal and background processes. The line shape of $\Lambda$(1405) varies with the four momentum transfer, $\mathrm{Q}^2$, and does not match the line shape based on PDG resonance parameters. It corresponds approximately to predictions of a recent two-pole meson-baryon picture of this state.
\end{abstract}
\maketitle


\section{Introduction}

The strong interaction, as one of the four fundamental interactions, is of central and enduring interest in particle physics. Baryons, as quantum systems of quarks ``glued'' by strong interaction, are a central subject to study strong interaction. In fact, Nathan Isgur asked ``Why N*'s'' in his closing speech at the workshop on Excited Nucleons and Hadronic Structure in 2000 \cite{proc:Isgur}.It was stated that  ``baryons were at the roots of the development of the quark model'' \cite{Klempt2010}. 

\begin{figure}
  \includegraphics[height=.22\textheight]{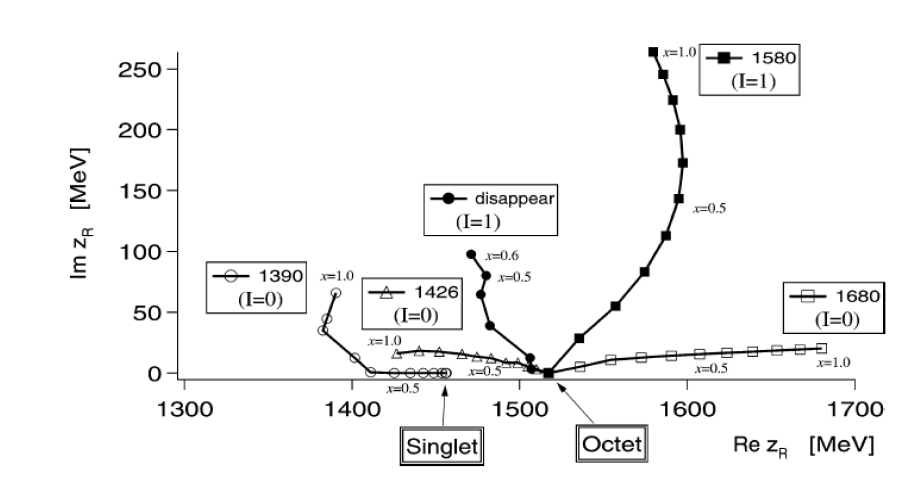}
  \caption{Prediction of the two pole positions given in \cite{Jido2003}. The pole positions starts at $x=0$ for exact SU(3) to $x=1$ for physical values. The two $I=0$ poles are predicted to form the observed $\Lambda$(1405). Figure taken from \cite{Jido2003}}\label{fig:jido2003}
\end{figure}

It was first predicted and proposed by \citet{Dalitz1959,Dalitz1960} that there is a bound-state resonance below $\bar{K}N$ threshold. It was first observed by \citet{Alston1961}. Following an earlier paper by \citet{Oller2001}, \citet{Jido2003} used a chiral unitary formalism to describe the $\Lambda$(1405) as formed by two $I=0$ poles, 1390+i0.066\ GeV and 1420+i0.016\ GeV. Figure \ref{fig:jido2003} illustrates the evolution of these poles from the exact SU(3) symmetry limit (x=0) to the physical mass limit (x=1), and shows how the 1405 MeV region is a composite of two states. The key prediction of this model was that the mass distribution, or ``lineshape'' of the $\Lambda$(1405) should depend upon which reaction channel excites it and into which final state it decays. No specific prediction was made for electroproduction of this state, so the present study represents an exploration of an unknown corner of $\Lambda$(1405) phenomenology.

\section{Event Selection and Background Subtraction}

\begin{figure}
  \includegraphics[height=.22\textheight]{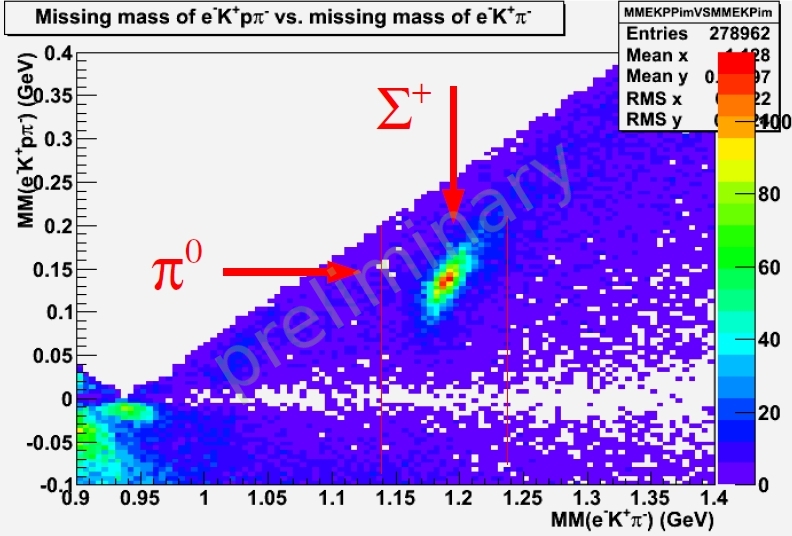}
  \caption{Missing mass of $e^-K^+p\pi^-$ vs. missing mass of $e^-K^+\pi^-$ from data.}\label{fig:PizVSSigma}
\end{figure}

This work analyzed the data from the E1F run in Hall B at Jefferson Lab, which utilized an electron beam of 5.5\ GeV and liquid hydrogen target.
The $\Lambda$(1405) decays into $\Sigma\pi$ 100\% according to PDG \citep{Nakamura2010}. The decay channel chosen in this work was $\Sigma^+\pi^-$ for the $\Lambda$(1405) and $p\pi^0$ for $\Sigma^+$. Therefore, the final detected particles were the scattered electron $e^-$, $K^+$, proton, and $\pi^-$, while a $\pi^0$ was missing for the exclusive electroproduction of $\Lambda$(1405). Thus, events with exactly these four charged particles were selected. 
The missing mass of $e^-K^+p\pi^-$ vs. the missing mass of $e^-K^+\pi^-$ is shown in Fig. \ref{fig:PizVSSigma}. The missing $\pi^0$s and $\Sigma^+$s are clearly seen as indicated in the figure. A selection was made by requiring the missing mass of $e^-K^+\pi^-$ between 1.14\ GeV and 1.24\ GeV and the missing mass of $e^-K^+p\pi^-$ between 0.05\ GeV and 0.2\ GeV. 
The $\mathrm{Q}^2$ range, after the selection, is mainly from 1.0\ $\mathrm{GeV}^2$ to 3.0\ $\mathrm{GeV}^2$.

\begin{figure}
  \includegraphics[height=.22\textheight]{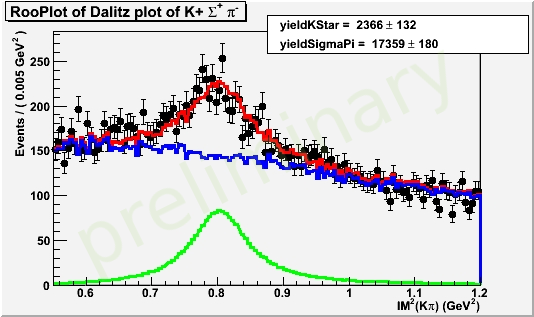}
  \caption{Invariant mass of $K^+\pi^-$. Points with error bars are the data, the matching line is fit to the data with two background simulations. The (green) line at the bottom is the simulation of $K^{*0}\Sigma^+$, and the upper (blue) line is simulation of $K^+\Sigma^+\pi^-$, while the uppermost (red) line is the sum.}\label{fig:FitRatio}
\end{figure}

The background is mostly from the nonresonant electroproduction of $K^+\Sigma^+\pi^-$ and resonant $K^{*0}\Sigma^+$, whose $K^{*0}$ decays into $K^+$ and $\pi^-$. The invariant mass of $K^+$ and $\pi^-$ was plotted and fitted with simulations of two channels. The total fit, with contribution from $K^{*0}\Sigma^+$ and $K^+\Sigma^+\pi^-$ is shown in Fig. \ref{fig:FitRatio}.

\section{Acceptance correction and $\mathrm{Q}^2$ dependence}

\begin{figure}
  \captionsetup{type=figure}
  \centering
  \subfloat[$\mathrm{Q}^2$ from 1.0 to 1.5\ $\mathrm{GeV}^2$]{\label{fig:accYield_lowQ2}\includegraphics[height=.20\textheight]{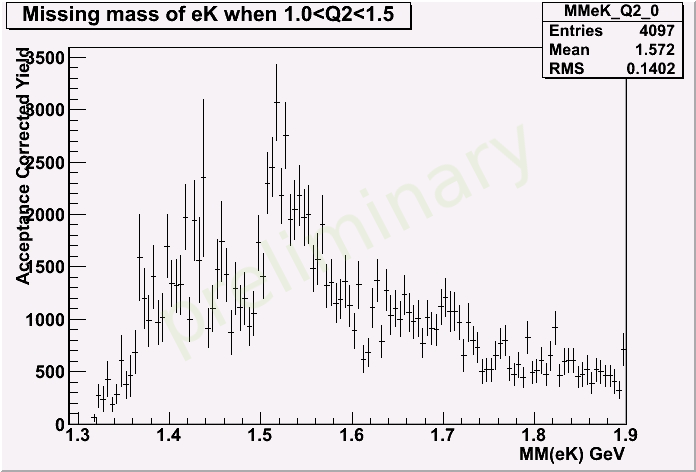}}
  \subfloat[$\mathrm{Q}^2$ from 1.5 to 3.0\ $\mathrm{GeV}^2$]{\label{fig:accYield_highQ2}\includegraphics[height=.20\textheight]{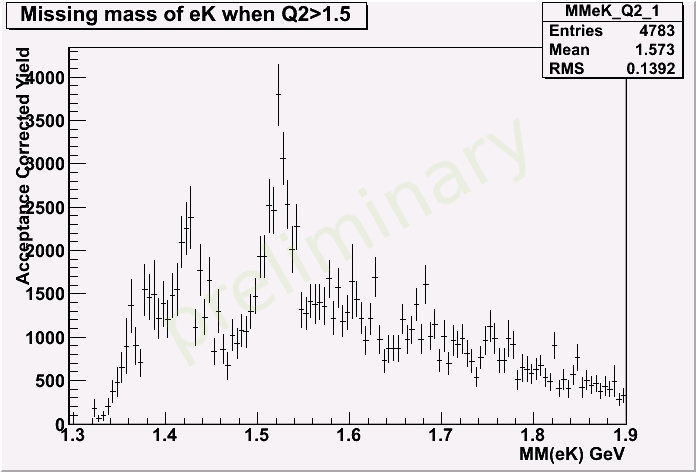}}
  \caption{Acceptance-corrected missing mass of $e^-K^+$. Two ranges of Q$^2$ are shown in (a) and (b).}\label{fig:accYield}
\end{figure}

\begin{figure}
  \includegraphics[height=.22\textheight]{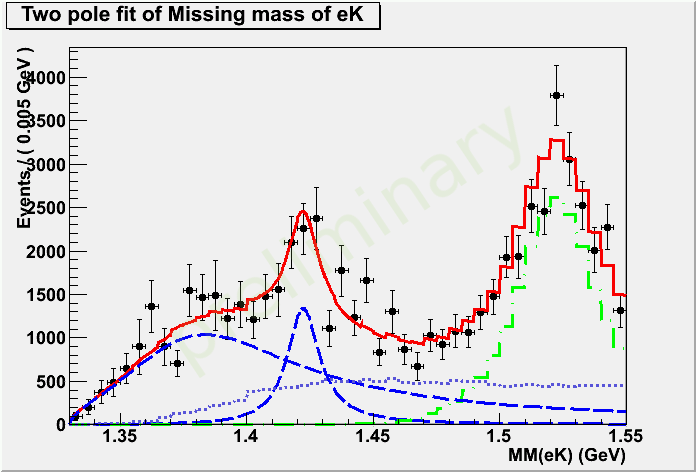}
  \caption{Fitting of the missing mass of $e^-K^+$. Points with error bars are the data, solid (red) line from overall fit, dash-dotted (green) line around 1.52\ GeV from $\Lambda$(1520) simulation, two dashed (blue) lines from relativistic Breit-Wigner function with fitted amplitude. The dotted (purple) line shows the summed background contributions.}
\label{fig:twopoleFit}
\end{figure}
Acceptance correction was performed using the simulation of non-resonant three-body phase space $K^+\Sigma^+\pi^-$. The raw and accepted events were binned in $\mathrm{Q}^2$ (1.0\ to 3.0\ $\mathrm{GeV}^2$), W (1.5 to 3.5 GeV) and cosine of the kaon angle in the center-of-mass frame. Acceptance factor was calculated for each of 9500 bins. Data was corrected bin by bin in order to reduce bin-averaging effects.
The corrected yield is shown in Fig. \ref{fig:accYield}, where Fig. \ref{fig:accYield}\subref{fig:accYield_lowQ2} is for $\mathrm{Q}^2$ from 1.0 to 1.5\ $\mathrm{GeV}^2$ and Fig. \ref{fig:accYield}\subref{fig:accYield_highQ2} for $\mathrm{Q}^2$ from 1.5 to 3.0\ $\mathrm{GeV}^2$. The statistics in these two Q$^2$ regions are similar to each other. The dominant peak in both figures is from the $\Lambda$(1520). A clear peak around 1.42\ GeV and another peak with much broader width at the lower invariant mass in Fig. \ref{fig:accYield}\subref{fig:accYield_highQ2} are not consistent with PDG values of $\Lambda$(1405) \cite{Nakamura2010}. However, they are not clearly seen in the lower Q$^2$ range ( Fig. \ref{fig:accYield}\subref{fig:accYield_lowQ2}). 
The higher Q$^2$ range distribution was fitted with simulated background, $\Lambda$(1520) and two relativistic Breit-Wigner functions described in Eqn. \ref{eqn:rbw}. 
\begin{equation}\label{eqn:rbw}
BW(m) \sim \frac{1}{2\pi}\frac{4mm_0\Gamma(q)}{(m^2-m^2_0)^2+(m_0\Gamma(q))^2}
\end{equation}
Here $\Gamma(q)=\frac{q}{q_0}\Gamma_0$, $m_0$ and $\Gamma_0$ are fitting parameters, $q$ is the momentum of $\pi^-$ in center-of-mass frame of $\Sigma^+\pi^-$, and $q_0$ is the value of $q$ when $m=m_0$.
The result gives the preliminary fit parameters as $m_0=1.422$\ GeV and $\Gamma_0=0.016$\ GeV for peak around 1.42\ GeV, and $m_0=1.393$\ GeV and $\Gamma_0=0.100$\ GeV for peak below 1.4\ GeV (Fig. \ref{fig:twopoleFit}). More detailed study of fits to this spectrum are in progress.

\section{Summary}

The $\Lambda$(1405) mass distribution in electroproduction is different from a single Breit-Wigner distribution and varies with Q$^2$. A fit with two relativistic Breit-Wigner functions has been performed and the fit result favors the two-pole theoretic prediction.

\begin{theacknowledgments}
  Author H. Lu would like to thank Kijun Park and Ralf Gothe for discussion and help.
\end{theacknowledgments}



\bibliographystyle{aipproc}   

\bibliography{nstar2011}

\hyphenation{Post-Script Sprin-ger}
\begin{thebibliography}{8}
\expandafter\ifx\csname natexlab\endcsname\relax\def\natexlab#1{#1}\fi
\providecommand{\enquote}[1]{``#1''}
\expandafter\ifx\csname url\endcsname\relax
  \def\url#1{\texttt{#1}}\fi
\expandafter\ifx\csname urlprefix\endcsname\relax\def\urlprefix{URL }\fi
\providecommand{\eprint}[2][]{\url{#2}}

\bibitem[Isgur(2001)]{proc:Isgur}
N.~Isgur, \enquote{Why N*'s Are Important,} in \emph{Excited Nucleons and
  Hadronic Structure, N*2000, Newport News, 2000}, edited by V.~D. Burkert,
  L.~Elonadrihiri, and J.~J. Kelly, World Scientific, Singapore, 2001, p. 403.

\bibitem[Klempt and Richard(2010)]{Klempt2010}
E.~Klempt, and J.-M. Richard, \emph{Rev. Mod. Phys.} \textbf{82}, 1095--1153
  (2010).

\bibitem[Dalitz and Tuan(1959)]{Dalitz1959}
R.~H. Dalitz, and S.~F. Tuan, \emph{Phys. Rev. Lett} \textbf{2}, 425 -- 428
  (1959).

\bibitem[Dalitz and Tuan(1960)]{Dalitz1960}
R.~H. Dalitz, and S.~F. Tuan, \emph{Ann. of Phys.} \textbf{10}, 307--351
  (1960).

\bibitem[Alston and et~al.(1961)]{Alston1961}
M.~H. Alston, and et~al., \emph{Phys. Rev. Lett.} \textbf{6}, 698--702 (1961).

\bibitem[Oller and Meissner(2001)]{Oller2001}
J.~A. Oller, and U.-G. Meissner, \emph{Physics Letters B} \textbf{500}, 263 --
  272 (2001).

\bibitem[Jido and et~al.(2003)]{Jido2003}
D.~Jido, and et~al., \emph{Nucl. Phys. A} \textbf{725}, 181 -- 200 (2003).

\bibitem[Nakamura and et~al.(2010)]{Nakamura2010}
K.~Nakamura, and et~al., \emph{J. Phys. G} \textbf{37}, 075021 (2010).

\end{thebibliography}

\IfFileExists{\jobname.bbl}{}
 {\typeout{}
  \typeout{******************************************}
  \typeout{** Please run "bibtex \jobname" to optain}
  \typeout{** the bibliography and then re-run LaTeX}
  \typeout{** twice to fix the references!}
  \typeout{******************************************}
  \typeout{}
 }

\end{document}